\documentclass{emulateapj}

\shorttitle{NuSTAR Observations of Tycho}
\shortauthors{LOPEZ ET AL.}

\newcommand{\ltsima}{$\; \buildrel < \over \sim \;$}
\newcommand{\simlt}{\lower.5ex\hbox{\ltsima}}

\newcommand{\ls}{{_<\atop^{\sim}}}

\newcommand{\gs}{{_>\atop^{\sim}}}

\def\arcmin{\hbox{$^\prime$}}
\def\arcsec{\hbox{$^{\prime\prime}$}}

\begin{document}

\title{A Spatially Resolved Study of the Synchrotron Emission and Titanium in Tycho's Supernova Remnant using NuSTAR}

\author{Laura A. Lopez\altaffilmark{1}, Brian W. Grefenstette\altaffilmark{2}, Stephen P. Reynolds\altaffilmark{3}, Hongjun An\altaffilmark{4}, Steven E. Boggs\altaffilmark{5}, Finn E. Christensen\altaffilmark{6}, William W. Craig\altaffilmark{5}, Kristoffer A. Eriksen\altaffilmark{7}, Chris L. Fryer\altaffilmark{7}, Charles J. Hailey\altaffilmark{8}, Fiona A. Harrison\altaffilmark{2}, Kristin K. Madsen\altaffilmark{2}, Daniel K. Stern\altaffilmark{9}, William W. Zhang\altaffilmark{10}, Andreas Zoglauer\altaffilmark{5}}
\altaffiltext{1}{Department of Astronomy and Center for Cosmology \& Astro-Particle Physics, The Ohio State University, Columbus, Ohio 43210}
\altaffiltext{2}{Cahill Center for Astrophysics, 1216 E. California Blvd., California Institute of Technology, Pasadena, CA 91125, USA}
\altaffiltext{3}{Physics Department, NC State University, Raleigh, NC 27695, USA}
\altaffiltext{4}{Department of Physics, McGill University, Montreal, Quebec, H3A 2T8, Canada}
\altaffiltext{5}{Space Sciences Laboratory, University of California, Berkeley, CA 94720, USA}
\altaffiltext{6}{DTU Space, National Space Institute, Technical University of Denmark, Elektrovej 327, DK-2800 Lyngby, Denmark}
\altaffiltext{7}{CCS-2, Los Alamos National Laboratory, Los Alamos, NM 87545, USA}
\altaffiltext{8}{Columbia Astrophysics Laboratory, Columbia University, New York, NY 10027, USA}
\altaffiltext{9}{Jet Propulsion Laboratory, California Institute of Technology, Pasadena, CA 91109, USA}
\altaffiltext{10}{NASA Goddard Space Flight Center, Greenbelt, MD 20771, USA}

\email{lopez.513@osu.edu}

\begin{abstract}

We report results from deep observations ($\sim$750 ks) of Tycho's supernova remnant (SNR) with {\it NuSTAR}. Using these data, we produce narrow-band images over several energy bands to identify the regions producing the hardest X-rays and to search for radioactive decay line emission from $^{44}$Ti. We find that the hardest ($>$10 keV) X-rays are concentrated in the southwest of Tycho, where recent {\it Chandra} observations have revealed high emissivity ``stripes'' associated with particles accelerated to the knee of the cosmic-ray spectrum. We do not find evidence of $^{44}$Ti, and we set limits on its presence and distribution within the SNR. These limits correspond to a upper-limit $^{44}$Ti mass of $M_{44} < 2.4 \times 10^{-4} M_{\sun}$ for a distance of 2.3~kpc. We perform spatially resolved spectroscopic analysis of sixty-six regions across Tycho. We map the best-fit rolloff frequency of the hard X-ray spectra, and we compare these results to measurements of the shock expansion and ambient density. We find that the highest energy electrons are accelerated at the lowest densities and in the fastest shocks, with a steep dependence of the roll-off frequency with shock velocity. Such a dependence is predicted by models where the maximum energy of accelerated electrons is limited by the age of the SNR rather than by synchrotron losses, but this scenario requires far lower magnetic field strengths than those derived from observations in Tycho. One way to reconcile these discrepant findings is through shock obliquity effects, and future observational work is necessary to explore the role of obliquity in the particle acceleration process.

\end{abstract} 


\section{Introduction}

Tycho's supernova remnant (SNR G120.1$+$1.4; hereafter Tycho) is widely believed to be the remnant of a Type Ia SN explosion observed in 1572 \citep{steph02,bad06,rest08,krause08}. Evidence of particle acceleration in Tycho was found first by \cite{pravdo79}, who detected non-thermal X-ray emission up to 25 keV with {\it HEAO-1}. Initial {\it Chandra X-ray Observatory} images spatially resolved the non-thermal X-ray emission and showed it originates from narrow filaments around the rim of Tycho \citep{hwang02,bamba05,warren05}. Such features are now thought to be common among young SNRs (see \citealt{reynolds08} for a review). A deep {\it Chandra} program revealed several non-thermal, high emissivity ``stripes'' in the projected interior of the SNR, a result which was interpreted as direct evidence of particles accelerated to the ``knee'' of the cosmic-ray (CR) spectrum at around 3 PeV (\citealt{eriksen11}, although see \citealt{bykov11} for an alternative interpretation). Recently, Tycho has been detected in GeV $\gamma$-rays with {\it Fermi} \citep{giordano12} and in TeV $\gamma$-rays with VERITAS \citep{acc11}. The $\gamma$-ray spectrum is consistent with diffusive-shock acceleration (DSA) either of CR protons \citep{berez13} or of two lepton populations \citep{atoyan12}. 

Recent X-ray studies of Tycho have reported the possible detection of radioactive decay lines of $^{44}$Ti \citep{troja14,wang14} and of Ti-K line emission at $\sim$4.9~keV \citep{miceli15}. As the properties of the $^{44}$Ti (like yield, spatial distribution and velocities) in young SNRs probe directly the underlying explosion mechanism of the supernova (see e.g., \citealt{mag10}), constraints on these parameters are useful to motivate and test simulations. The $^{44}$Ti isotope has a half-life of 60~years \citep{ahmad06} and originates from $\alpha$-rich freeze-out near nuclear statistical equilibrium (e.g., \citealt{timmes96,woosley02}). Radioactive decay lines of $^{44}$Ti to $^{44}$Sc and then to $^{44}$Ca are observable at $\sim$68, 78, and 1157 keV, and these features have been detected in young core-collapse SNRs Cassiopeia~A (e.g., \citealt{iyudin94,vink01,renaud06,brian14}) and SN~1987A \citep{grebenev12,boggs15}. $^{44}$Ti can also manifest itself through a 4.1~keV line which occurs from the filling of the inner-shell vacancy resulting from the decay of $^{44}$Ti to $^{44}$Sc by electron capture. A feature at this energy has been detected in {\it Chandra} X-ray spectra of the young SNR~G1.9$+$0.3 \citep{bork10,bork11}. To date, no Type Ia SNRs yet have definitive detections of the 68, 78, or 1157 keV lines associated with $^{44}$Ti (e.g., \citealt{dupraz97,iyudin99,renaud06b,zog14}). 

In this paper, we present hard X-ray images and spatially resolved spectra from a set of {\it NuSTAR} observations of Tycho totaling $\sim$750~ks. Launched in 2012, {\it NuSTAR} is the first satellite to focus at hard X-ray energies of 3--79 keV \citep{harrison13}. The primary scientific motivation of our observing program was to map and characterize the non-thermal X-ray emission in Tycho and to detect or constrain the spatial distribution of $^{44}$Ti. The paper is outlined as follows. In Section~\ref{sec:data}, we describe the data reduction and analysis procedures. In Section~\ref{sec:images}, we present the composite narrow-band {\it NuSTAR} images of the hard X-ray emission in Tycho, and in Section~\ref{sec:ti}, we exploit {\it NuSTAR} images to set upper limits on the presence of $^{44}$Ti. In Section~\ref{sec:spectra}, we report the results from a systematic spatially resolved spectroscopic analysis across the SNR. In Section~\ref{sec:discuss}, we discuss the implications of our results, specifically related to the particle acceleration properties of Tycho (in Section~\ref{sec:discuss_pa}) and to $^{44}$Ti searches in young SNRs (in Section~\ref{sec:discuss_ti}). We summarize our conclusions in Section~\ref{sec:conclusions}.

\section{Observations and Data Analysis} \label{sec:data}

Tycho was observed by {\it NuSTAR} three times from April--July 2014, as listed in Table~\ref{table:obslog}, with a total net integration of 748 ks. We reduced these data using the {\it NuSTAR} Data Analysis Software (NuSTARDAS) Version 1.3.1 and {\it NuSTAR} CALDB Version 20131223. We performed the standard pipeline data processing with {\it nupipeline}, with the stricter criteria for the passages through the South Atlantic Anomaly (SAA) and the ``tentacle''-like region near the SAA to reduce background uncertainties. 

\begin{deluxetable}{lrr}
\tablecolumns{3}
\tablewidth{0pt} \tablecaption{{\it NuSTAR} Observation Log \label{table:obslog}} 
\tablehead{\colhead{ObsID} & \colhead{Exposure} & \colhead{UT Start Date}}
\startdata
40020001002 & 339~ks & 2014 April 12 \\ 
40020011002 & 147~ks &  2014 May 31 \\
40020001004 & 262~ks & 2014 July 18
\enddata
\end{deluxetable}

Using the resulting cleaned event files, we produced images of different energy bands using the FTOOL {\it xselect} and generated associated exposure maps using {\it nuexpomap}. As Tycho is a bright, extended source, we opted to model the background and produce synthetic, energy-dependent background images for background subtraction. We followed the procedure detailed in \cite{wik14} and \cite{brian14} to estimate the background components and their spatial distribution. Subsequently, we combined the vignetting- and exposure-corrected FPMA and FPMB images from all epochs using {\it ximage}.

\begin{figure}
\begin{center}
\includegraphics[width=\columnwidth]{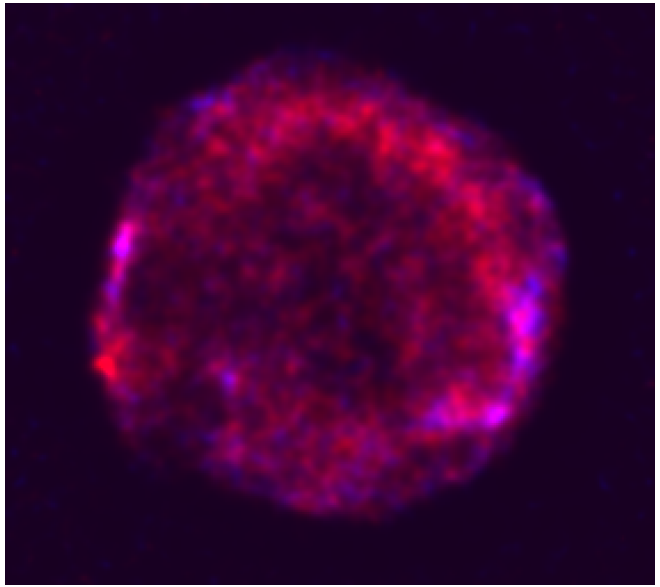}
\end{center}
\caption{Deconvolved, background-subtracted {\it NuSTAR} image from the combined $\sim$750~ks observation of Tycho, where red is 6--7 keV (which is dominated by Fe line emission) and blue is 10--20 keV. The SNR is $\sim$9\arcmin\  in diameter. In this and all subsequent images, north is up, and east is left.}
 \label{fig:twocolor}
\end{figure}

\begin{figure*}
\includegraphics[width=0.32\textwidth]{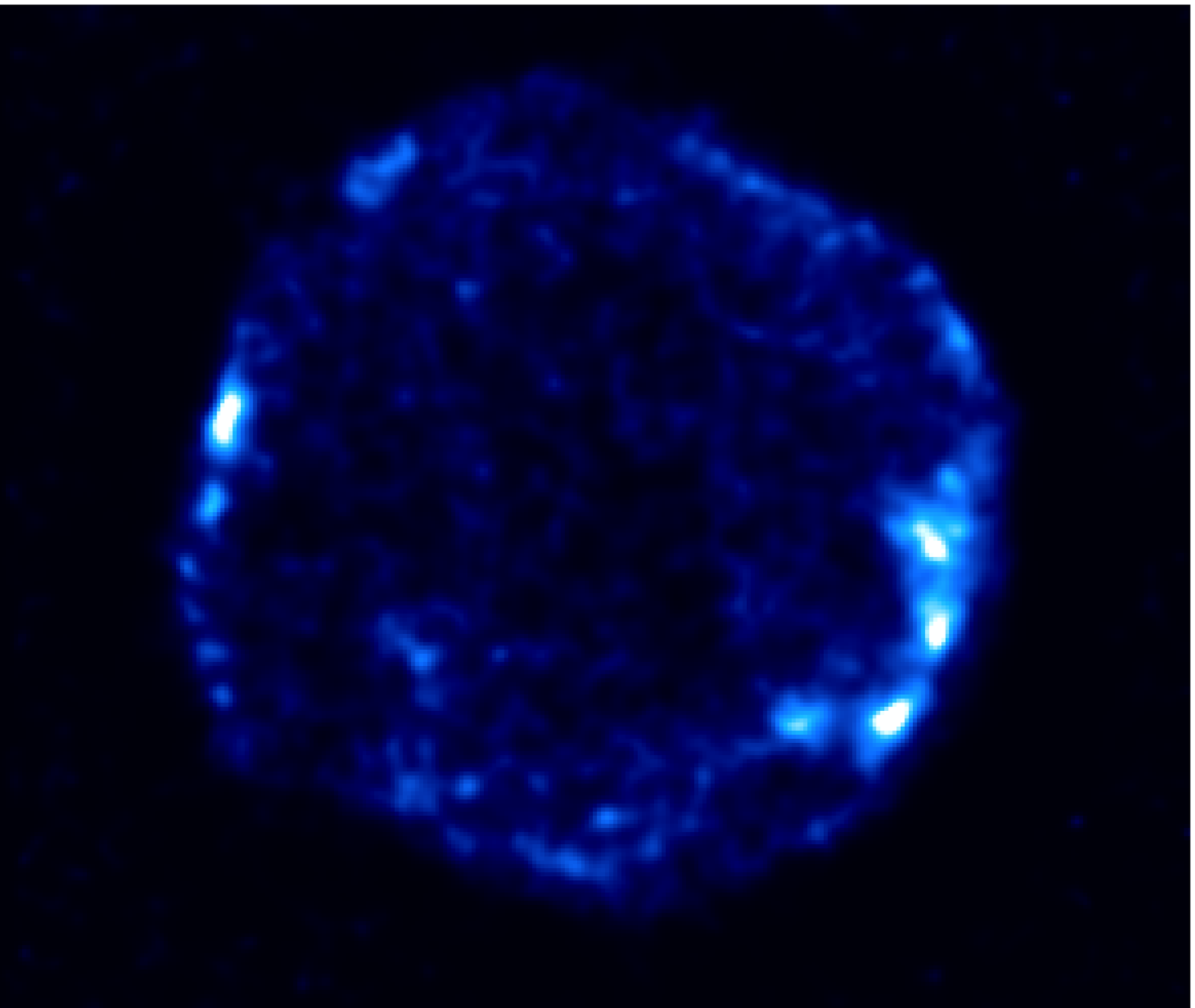}
\includegraphics[width=0.32\textwidth]{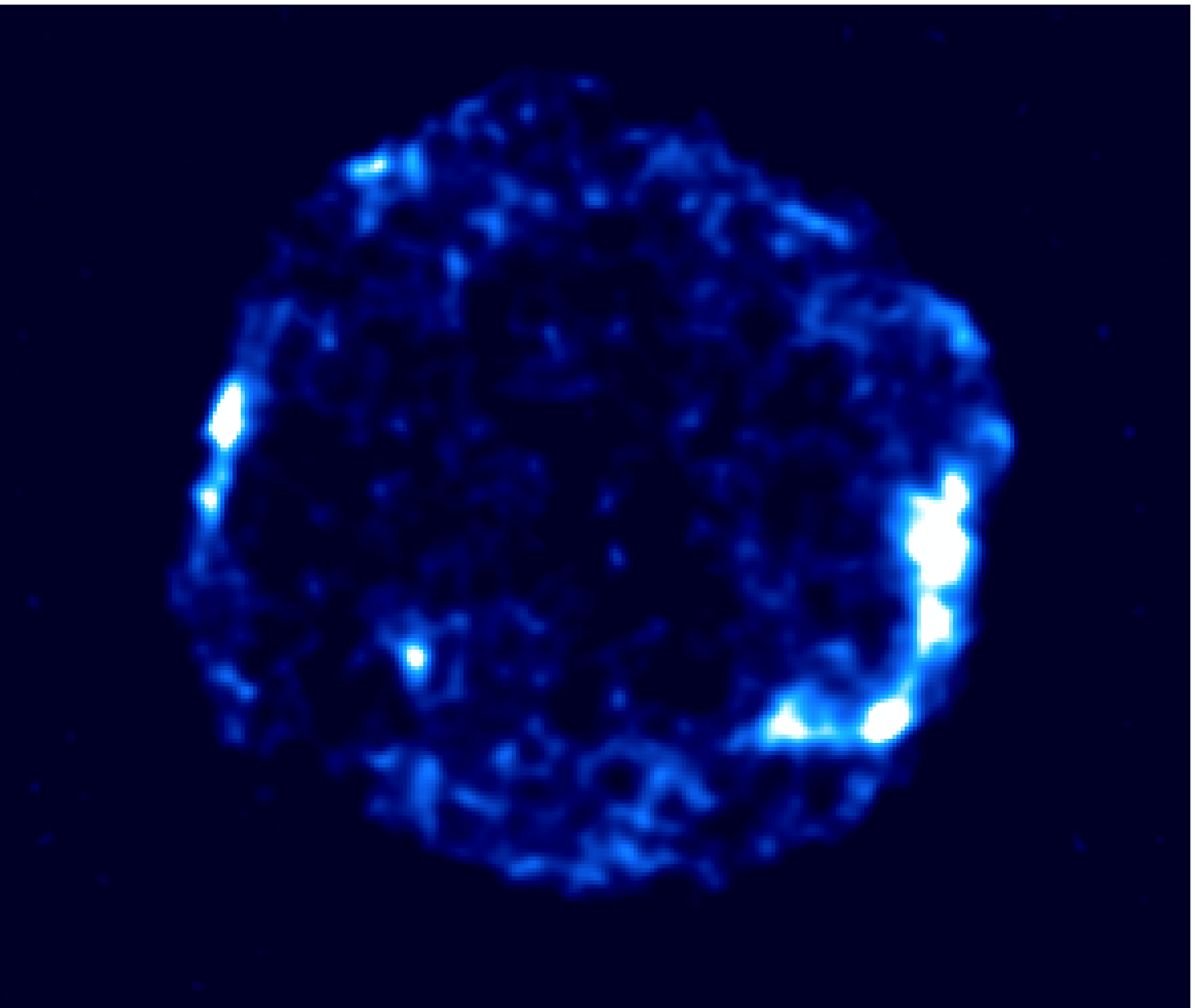}
\includegraphics[width=0.32\textwidth]{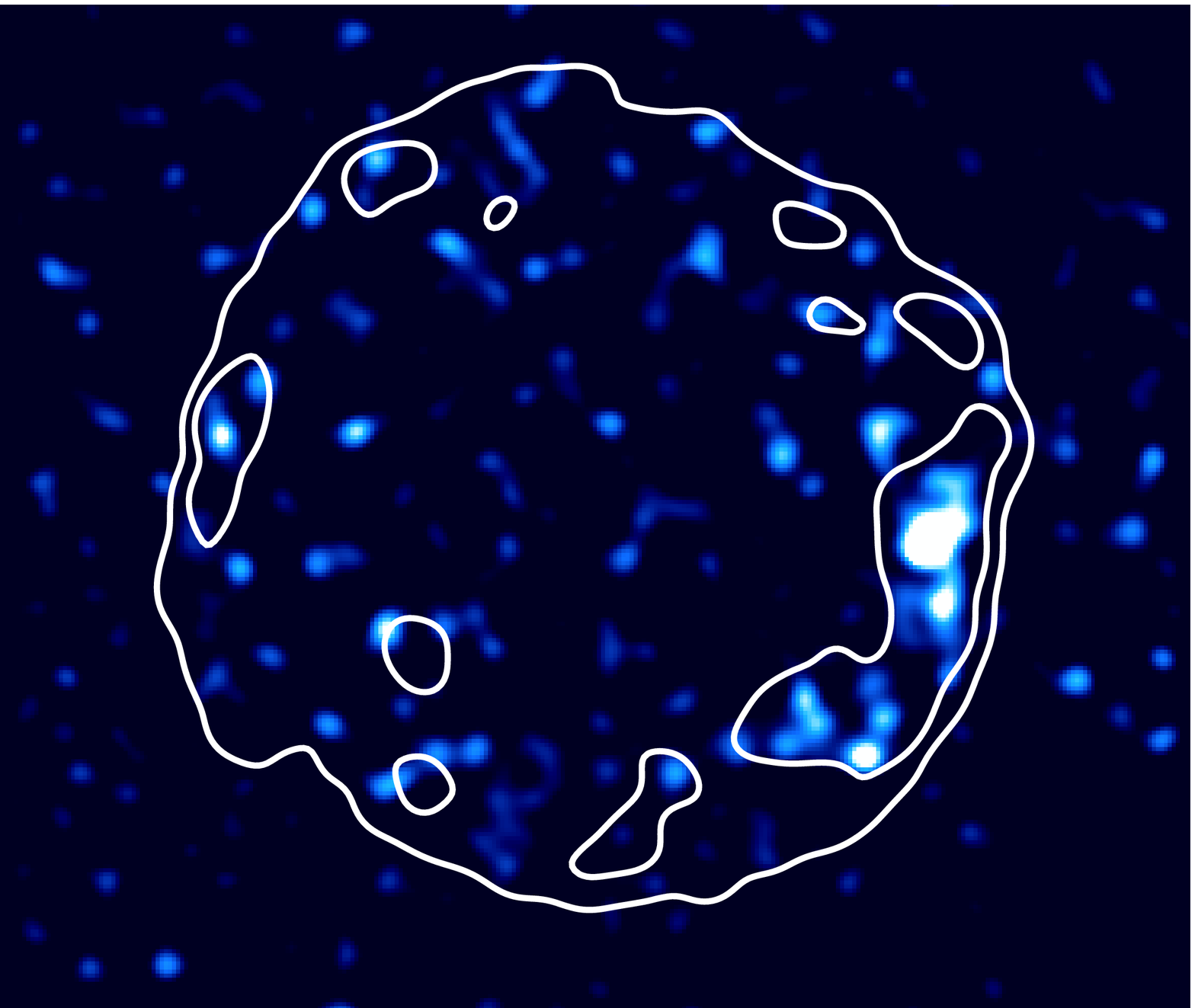}
\caption{Deconvolved, background-subtracted {\it NuSTAR} images of three hard energy bands: 8--10 keV (left), 10--20 keV (middle), and 20--40 keV (right). White contours on right panel are from the 10--20 keV band to guide the eye.}
 \label{fig:nt}
\end{figure*}

The combined images were then deconvolved by the on-axis {\it NuSTAR} point-spread function (PSF) using the {\it max\_likelihood} AstroLib IDL routine. The script employs Lucy-Richardson deconvolution, an iterative procedure to derive the maximum likelihood solution. We set the maximum number of iterations to 50, as more iterations did not lead to any significant changes in the resulting images. We note that this routine assumes the data can be characterized by a Poisson distribution, but background subtraction causes the images to not follow strictly a Poisson distribution. Thus, the deconvolved images in Figures~\ref{fig:twocolor} and ~\ref{fig:nt} are presented for qualitative purposes only, and we do not use them for any quantitative results.

We performed spatially resolved spectroscopic analyses by extracting and modeling spectra from locations across the SNR. Using the {\it nuproducts} FTOOL, we extracted source spectra and produced ancillary response files (ARFs) and redistribution matrix files (RMFs) from each of the three observations and both the A and B modules (3 ObsIDS $\times$ 2 modules = 6 spectra per region). Furthermore, we utilized the {\it nuskybgd} routines (presented in detail in \citealt{wik14}) to simulate the associated background spectra. We employed the {\it addascaspec} FTOOL to combine the six source pulse-height amplitude (PHA) files, ARFs, and background PHA files associated with each region. The six RMFs were added using the {\it addrmf} FTOOL. As discussed by \cite{gref15}, the addition of spectra from different observations leads to systematic effects of order a few percent in the best-fit normalizations of the spectra. 

\begin{figure*}
\includegraphics[width=\textwidth]{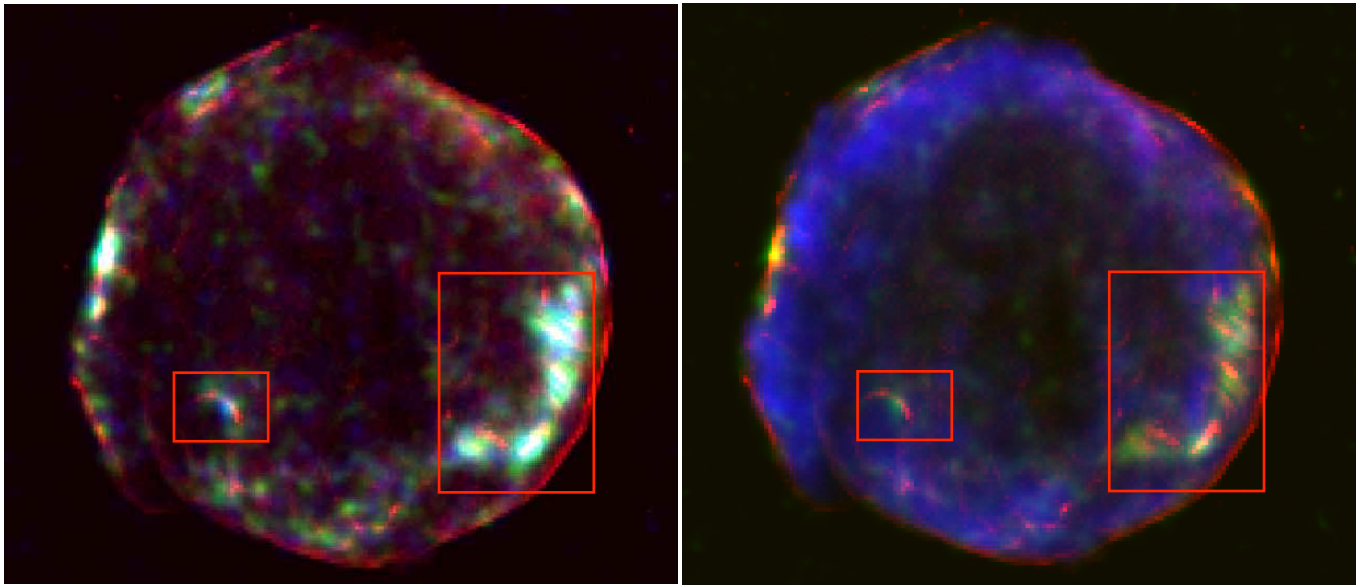}
\caption{{\it Left}: Three-color image, with 4--6 keV {\it Chandra} data in red (binned by a factor of eight), 8--10 keV deconvolved {\it NuSTAR} data in green, and 10--20 keV deconvolved {\it NuSTAR} data in blue. The hardest X-rays are brightest in the west of the SNR and coincide with the non-thermal X-ray ``stripes'' (denoted by the large red rectangle) seen in the deep {\it Chandra} data and analyzed by \cite{eriksen11}. The small rectangle corresponds to the ``hook'' of non-thermal emission described in the text. {\it Right}: Three-color image, with 4--6 keV {\it Chandra} data in red, 10--20 keV deconvolved {\it NuSTAR} data in green, and 20-cm VLA data in blue. We note that the {\it NuSTAR}, {\it Chandra}, and VLA images have pixel sizes of 2.5\arcsec, 3.9\arcsec, and 4.2\arcsec, respectively.}
\label{fig:chandracompare}
\end{figure*}

For comparison to the {\it NuSTAR} data, we analyzed archival {\it Chandra} and National Radio Astronomy Observatory (NRAO) Very Large Array (VLA) observations of Tycho. The {\it Chandra} Advanced CCD Imaging Spectrometer (ACIS) observed Tycho in April 2009 for a total integration of 734~ks \citep{eriksen11}. Reprocessed data were downloaded from the {\it Chandra} archive (ObsIDs 10093--10097, 10902--10906), and composite exposure-corrected images were produced of the 4--6 keV continuum using the {\it flux\_image} command in the {\it Chandra} Interactive Analysis of Observations ({\sc ciao}) software Version 4.6. The VLA observed Tycho at 1.505~GHz on 10 July 2001 in the C configuration for 174 minutes  (program AR464, PI: Reynoso). The reduced image, with a beam size of 15.8\arcsec\ by 13.5\arcsec, was downloaded from the NRAO/VLA image archive\footnotetext{https://science.nrao.edu/facilities/vla/archive/index}.

\section{Results} 
\subsection{{\it NuSTAR} Imaging} \label{sec:images}

Figures~\ref{fig:twocolor} and \ref{fig:nt} show the deconvolved, background-subtracted {\it NuSTAR} images of Tycho from the combined $\sim$750~ks observation in several energy bands. {\it NuSTAR} detects the ``fluffy'' Fe-rich ejecta found predominantly in the northwest of Tycho \citep{decour01,yamaguchi14} as well as non-thermal X-rays with up to energies of $\sim$40 keV. The hard ($>$8 keV) X-rays are predominantly distributed around the rim of the SNR, with the peak at the location of the ``stripes'' identified in {\it Chandra} observations (see Figure~\ref{fig:chandracompare} for a comparison). 

In the right panel of Figure~\ref{fig:chandracompare} we compare the 20-cm radio morphology seen by the VLA to the 4--6 keV {\it Chandra} and 10--20 keV {\it NuSTAR} images. The radio emission is distributed in a ring interior to the narrow, non-thermal filaments at the SNR periphery found by {\it Chandra}. Generally, the bright, hard X-ray knots detected by {\it NuSTAR} do not coincide with the radio emission. Overall, we find that the {\it Chandra} 4--6 keV band has a similar morphology as revealed in the {\it NuSTAR} images up to $\sim$20 keV. The hardest X-rays (20--40 keV) are concentrated around the stripes, as well as near a ``hook'' in the southwest of the SNR (denoted by the small red rectangle in Figure~\ref{fig:chandracompare}). 

\begin{figure}
\begin{center}
\includegraphics[width=\columnwidth]{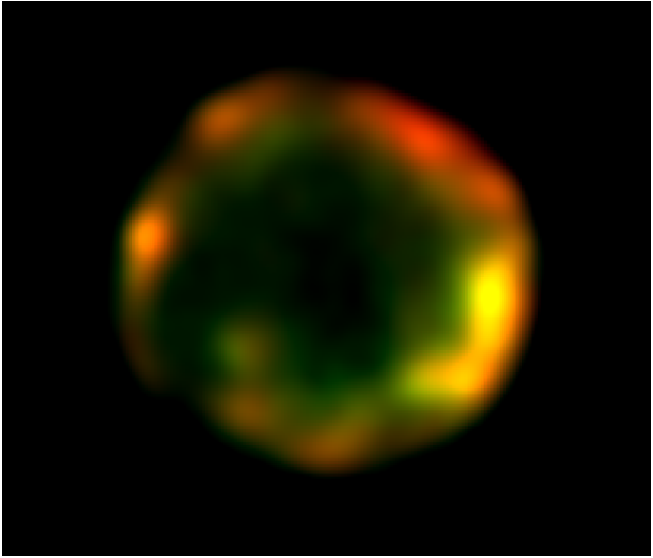}
\end{center}
\vspace{-3mm}
\caption{Two-color image of Tycho to demonstrate the spatial coincidence of the 4--6 keV {\it Chandra} data (in red) and the 10--20 keV deconvolved {\it NuSTAR} data (in green). We have convolved the {\it Chandra} image with a Gaussian function to match the resolution of the {\it NuSTAR} image. As the emission in the 10--20 keV band is almost exclusively non-thermal, locations where the 4--6 keV emission is coincident indicates that the latter is also non-thermal in nature.}
 \label{fig:identifynt}
\end{figure}

We also convolved the {\it Chandra} 4--6 keV image with a Gaussian to match the resolution of the {\it NuSTAR} data, and the comparison of the result to the 10--20 keV {\it NuSTAR} data is shown in Figure~\ref{fig:identifynt}. As the 10--20 keV emission is almost exclusively non-thermal, the locations where the 4--6 keV emission is coincident reveals the latter is substantially non-thermal in nature. The converse scenario (where 4--6 keV emission is found without 10--20 keV emission) indicates the softer band may be dominated by thermal emission. We find that the 4--6 keV emission is largely coincident with the harder band, except in the northwest, where previous X-ray studies have found significant emission from the ejecta \citep{decour01,yamaguchi14}. Thus, we conclude that the 4--6 keV {\it Chandra} band is likely dominated by non-thermal across Tycho except in the northwest.

\begin{figure}
\begin{center}
\includegraphics[width=\columnwidth]{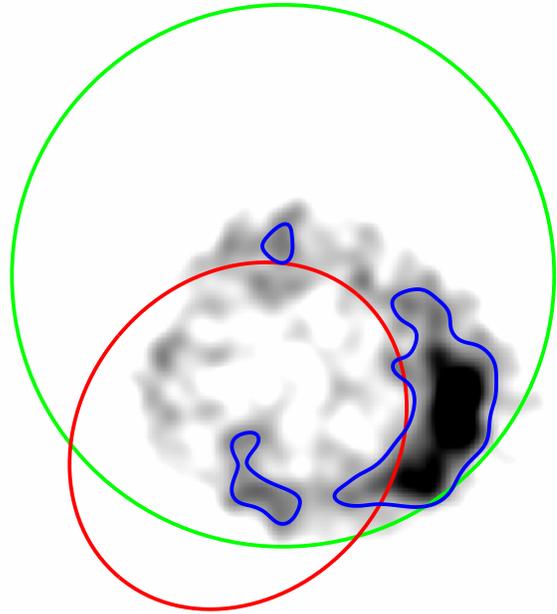}
\end{center}
\caption{Comparison of the high-energy detections of Tycho, with the {\it Fermi} 95\% confidence error circle denoted in red (for photons of energy $E > 1$ GeV; from \citealt{giordano12}), the VERITAS error circle in green (for energies $E>1$ TeV; from \citealt{acc11}), and the {\it NuSTAR} 20--40 keV band contours in blue. The image is the background-subtracted {\it NuSTAR} 20--40 keV image smoothed with a Gaussian function of width $\sigma =$ 5 pixels.}
\label{fig:highenergy}
\end{figure}

In Figure~\ref{fig:highenergy}, we compare the {\it NuSTAR} 20--40 keV emission to the GeV and TeV centroids reported by \cite{giordano12} and \cite{acc11}, respectively. The GeV localization (from {\it Fermi} observations with a 95\% confidence level) coincides with the hard X-rays from the ``hook'' feature in the southeast of the SNR, although we note that the ``stripe'' region is just outside the 95\% error circle. The TeV detection is consistent with a point source at the 0.11$^{\circ} \approx 6.6\arcmin$ PSF of VERITAS, which is of the same order as the angular extent  ($\sim$8\arcmin) of the SNR. Thus, it is not possible with the currently reported VERITAS observations to determine the origin of the TeV emission.

\subsection{Upper Limits on $^{44}$Ti} \label{sec:ti}

We do not find any point-like or extended emission in the remnant using the 65--70 keV band containing the 68 keV $^{44}$Ti emission line. We searched using the method from \cite{brian14}, where we compared the observed image to a simulated background image. We searched on multiple spatial scales by convolving both the {\it NuSTAR} image and the simulated image with a top hat function, varying the radius of the function from 3--15 {\it NuSTAR} pixels (roughly 7.5 to 30 arcseconds). We computed the Poisson probability that the observed counts in each pixel could have been produced by the predicted background flux, where a low probability indicates that the pixel likely contains a source count. We do not find any evidence for excess emission on any spatial scales.

\begin{figure*}
\begin{center}
\includegraphics[width=0.48\textwidth]{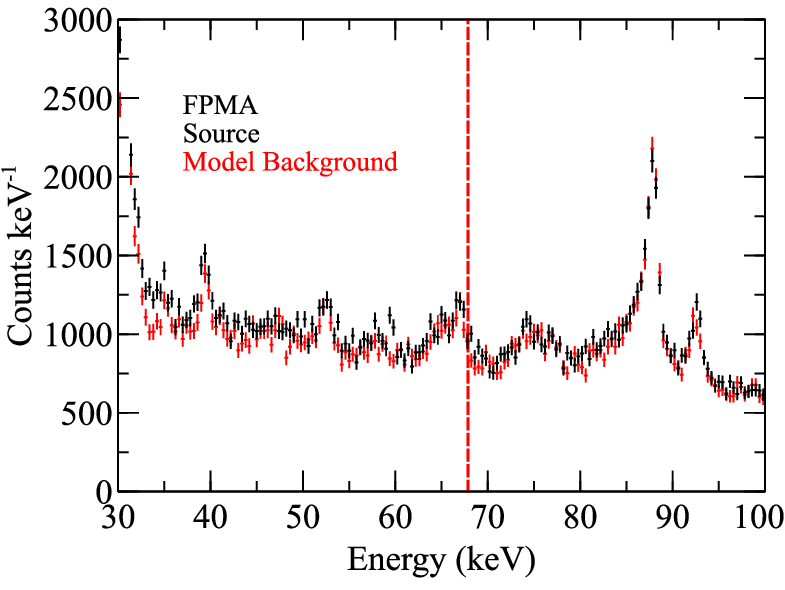}
\includegraphics[width=0.48\textwidth]{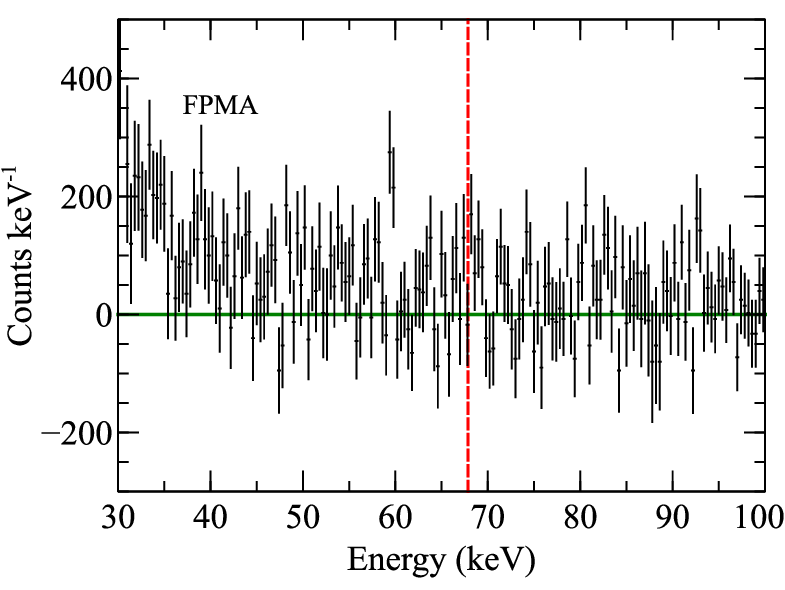}
\includegraphics[width=0.48\textwidth]{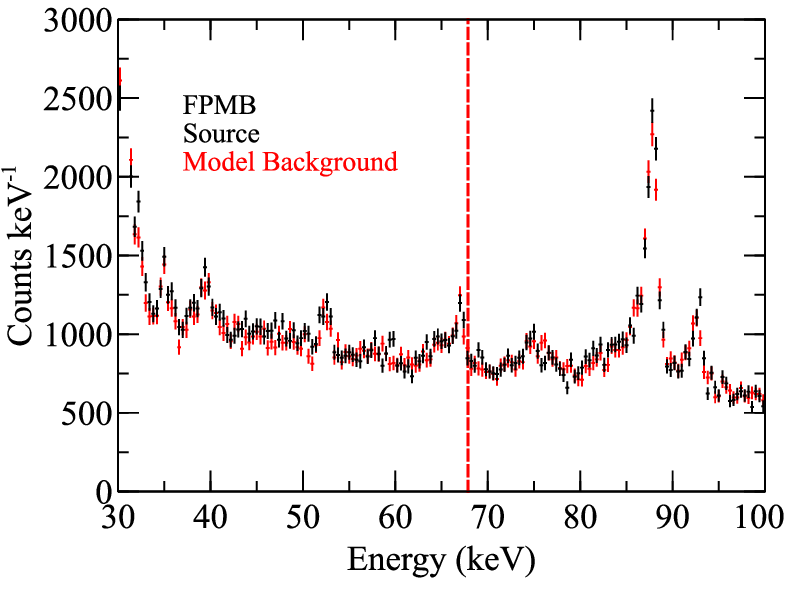}
\includegraphics[width=0.48\textwidth]{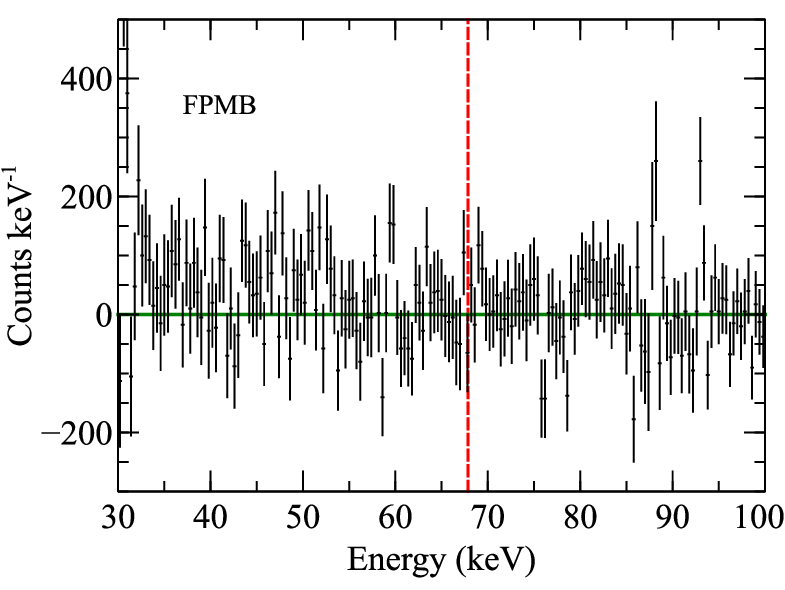}
\end{center}
\caption{{\it Top Left}: Integrated FPMA source spectra (black) and background model (red) of a 4\arcmin\ radius region. {\it Top right}: FPMA background-subtracted spectra. {\it Right panels}: Same as left panels for FPMB. Vertical red dashed line shows the rest-frame energy of the 67.86~keV Ti line. We do not find any excess emission at this line energy relative to the background model, and thus we set upper limits on $^{44}$Ti.}
\label{fig:examplespec}
\end{figure*}

We place upper limits on the diffuse extended emission in Tycho arising from the 68 keV line. We extracted spectra and produced an ARF and RMF for source regions with radii from 1 to 5 arcminutes centered on the remnant. For example, in Figure~\ref{fig:examplespec}, we give the spectra and background model of the FPMA and FPMB data for a 4\arcmin\ radius extraction region. For each radius, we modeled the source using a combination of a power-law continuum with two Gaussian lines of fixed relative intensity centered at 68 and 78 keV and fit the data over the 10--79 keV bandpass. We do not find any evidence for line emission in any of the source regions. 

\begin{figure}
\begin{center}
\includegraphics[width=\columnwidth]{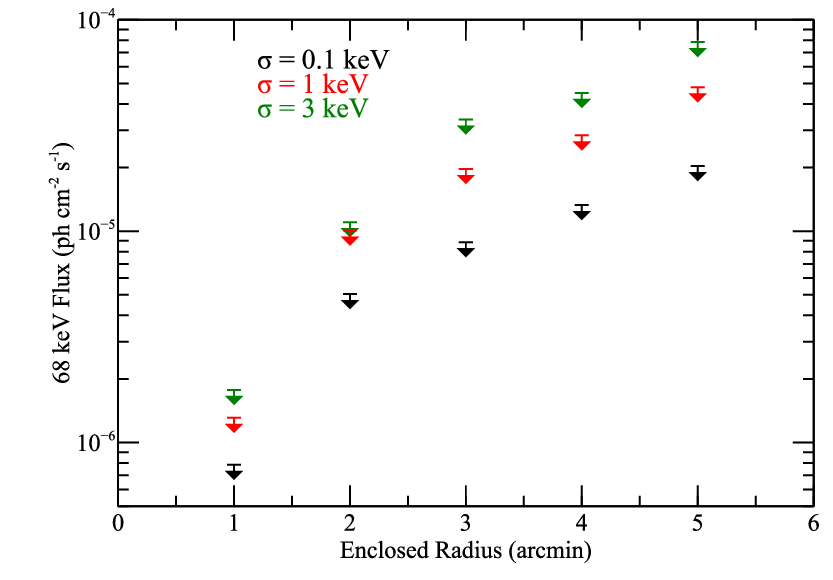}
\end{center}
\caption{The black, red, and green points show the 3-$\sigma$ confidence limits as a function of radius for line widths $\sigma$ of 0.1, 1, and 3 keV (corresponding to velocities of $\sim$730, 7300, and 22,000 km s$^{-1}$). }
\label{fig:limits}
\end{figure}

Our upper limits are dependent on the intrinsic width of the $^{44}$Ti lines. We computed the 3-$\sigma$ confidence limits on the line normalizations for fixed line widths of 0.1, 1, or 3 keV, correspond to velocities of $\approx$733, 7330, and 22,000 km~s$^{-1}$, respectively. The resulting 3-$\sigma$ upper limits as a function of radius are shown in Figure~\ref{fig:limits}. In all cases, the lower limits are zero. 

To effectively compare these {\it NuSTAR} upper limits with results from other missions, we convert all previous measurements and limits to epoch 2014. In Table~\ref{table:limits}, we list these values for observations with the {\it Compton} Gamma-ray Observatory/COMPTEL, {\it INTEGRAL}/IBIS, and {\it Swift}/BAT. To obtain epoch 2014 fluxes (which are also given in Table~\ref{table:limits}), we adopt a half-life of 58.9 years \citep{ahmad06}. The {\it NuSTAR} upper limits may be used in conjunction with the previous results to assess the likely spatial distribution of the $^{44}$Ti. In particular, our upper limits for radii $\ls$2\arcmin\ are less than the reported {\it Swift}/BAT detection (at epoch 2014) of (1.2$\pm$0.6)$\times10^{-5}$ ph cm$^{-2}$ s$^{-1}$, suggesting that the $^{44}$Ti is distributed over a larger radius of Tycho. 

Recently, \cite{miceli15} reported detection of the Ti-K line complex at $\sim$4.9~keV using stacked {\it XMM-Newton} spectra from Tycho. In particular, these authors identified the Ti feature in two regions of enhanced emission from Fe-group elements, in the northwest and in the southeast of the SNR (regions \#1 and \#3 in Figure~2 of \citealt{miceli15}). Motivated by this result, we searched for the 68~keV line at these locations in the {\it NuSTAR} data (the region files were provided graciously by M.~Miceli). We did not detect any signal above the background from either region, with lower limits of zero in all cases. The 3-$\sigma$ confidence flux upper limits are 3.9$\times10^{-6}~{\rm ph}~{\rm cm}^{-2}~{\rm s}^{-1}$ and 7.3$\times10^{-7}~{\rm ph}~{\rm cm}^{-2}~{\rm s}^{-1}$ for regions \#1 and \#3, respectively, assuming a line-width of 1 keV. 

\begin{deluxetable*}{lcccc}
\tablecolumns{5}
\tablewidth{0pt} \tablecaption{Previous Measurements and Limits\tablenotemark{a}  on $^{44}$Ti \label{table:limits}} 
\tablehead{\colhead{Instruments} & \colhead{Years of} & \colhead{Mean Year} & \colhead{Flux at Mean Year} & \colhead{Flux at Epoch 2014} \\
\colhead{} & \colhead{Obs.} &  \colhead{of Obs.} &  \colhead{($\times10^{-5}$ ph cm$^{-2}$ s$^{-1}$)} & \colhead{($\times10^{-5}$ ph cm$^{-2}$ s$^{-1}$)}}
\startdata
{\it CGRO}/COMPTEL & 1991--1997 & 1994 & $<$1.6 & $<$1.3 \\ 
{\it INTEGRAL}/IBIS & 2002--2006 & 2004 & $<$1.5 & $<$1.4 \\
{\it Swift}/BAT & 2004--2013 & 2008.5 & 1.3$\pm$0.6 & 1.2$\pm$0.6 \\
\enddata
\tablenotetext{a}{{\it CGRO}/COMPTEL upper limit is 2-$\sigma$, from \cite{dupraz97} and \cite{iyudin99}; {\it INTEGRAL}/IBIS upper limit is 3-$\sigma$, from \cite{renaud06b} and \cite{wang14}; the {\it Swift}/BAT detection is the 90\% confidence limit, from \cite{troja14}.}
\end{deluxetable*}

\subsection{Spatially Resolved Spectroscopy with {\it NuSTAR}} \label{sec:spectra}

\begin{figure}
\begin{center}
\includegraphics[width=\columnwidth]{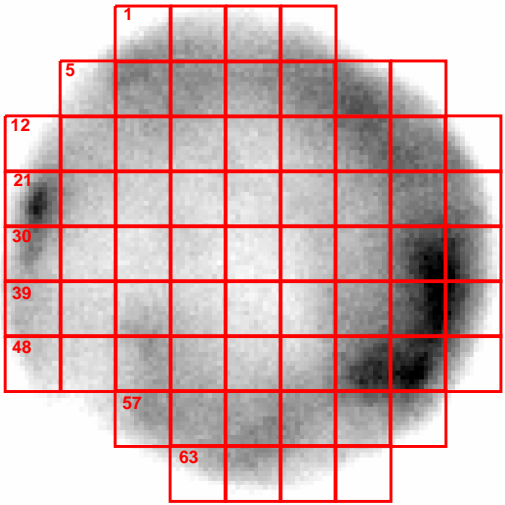}
\end{center}
\caption{Background-subtracted, merged 3--20 keV image of Tycho with 1\arcmin\ by 1\arcmin\ boxes overplotted denoting the regions where {\it NuSTAR} spectra were extracted and modeled.}
\label{fig:boxes}
\end{figure}

We performed spatially resolved spectroscopic analyses by extracting and modeling spectra from sixty-six 1\arcmin\ by 1\arcmin\ boxes across the SNR (see Figure~\ref{fig:boxes}). To evaluate the non-thermal emission, we fit the spectra in the 10--50 keV range with an absorbed {\it srcut} model. In our fits, we have fixed the absorbing column to $N_{\rm H} = 7 \times 10^{21}$ cm$^{-2}$ \citep{eriksen11}, and we have grouped the spectra to a minimum of 30 counts bin$^{-1}$. 

The {\it srcut} model describes the spectrum as that radiated by a power-law energy distribution of electrons with an exponential cutoff at energy $E_{\rm max}$ \citep{reynolds99}. That spectrum cuts off roughly as $\exp(-\sqrt{\nu/\nu_{\rm rolloff}})$ (though the fitting uses the complete expression). The rolloff photon energy $E_{\rm rolloff} = h \nu_{\rm rolloff}$ is related to the maximum energy of the accelerated electrons $E_{\rm max}$ by

\begin{equation}
E_{\rm max} = 120  \bigg(\frac{h \nu_{\rm rolloff}}{{\rm 1~keV}} \bigg)^{1/2} \bigg(\frac{B}{\mu{\rm G}} \bigg)^{-1/2}~~{\rm TeV}
\label{eq:Emax}
\end{equation}

\noindent
where $B$ is the magnetic field strength and we have corrected for a small numerical error in \cite{reynolds99}. 

In XSPEC, the {\it srcut} model is characterized by three parameters: the rolloff frequency $\nu_{\rm rolloff}$, the mean radio-to-X-ray spectral index $\alpha$, and the 1~GHz radio flux density $F_{\rm 1GHz}$. To limit the free parameters in the fit, we estimated $F_{\rm 1GHz}$ in each of the sixty-six regions by measuring the 1.505~GHz flux density in the NRAO/VLA data assuming a radio spectral index of $\alpha = -0.65$ \citep{kothes06}. The derived flux densities at 1~GHz using this procedure are listed in Table~\ref{tab:spec}. We note that without additional single-dish observations, the interferometric VLA data are missing flux from the largest scales. Based on previous radio studies of Tycho, we estimate the missing flux is of order $\sim$10\% of the values listed in Table~\ref{tab:spec}. 

\begin{deluxetable*}{lccc| lccc}
\tablecolumns{8}
\tablewidth{0pt} \tablecaption{Spatially Resolved Spectroscopy Fit Results Using SRCUT\tablenotemark{a} \label{tab:spec}}
\tablehead{\colhead{Region} & \colhead{$F_{\rm 1 GHz}$} & \colhead{$\nu_{\rm rolloff}$} & \colhead{$\chi^{2}$/dof} & \colhead{Region} & \colhead{$F_{\rm 1 GHz}$} & \colhead{$\nu_{\rm rolloff}$} & \colhead{$\chi^{2}$/d.o.f.} \\
\colhead{} & \colhead{(Jy)} & \colhead{($\times10^{17}$~Hz)} & \colhead{} & \colhead{} & \colhead{(Jy)} & \colhead{($\times10^{17}$~Hz)} & \colhead{}} 
\startdata
1 & 0.55 & 1.91$\pm$0.06 & 109/126 & 34 & 0.42 & 1.55$\pm$0.05 & 117/110 \\
2 & 1.32 & 1.17$\pm$0.03 & 166/138 & 35 & 0.53 & 1.59$\pm$0.05 & 138/123 \\
3 & 1.29 & 1.08$\pm$0.03 & 130/122 & 36 & 0.75 & 1.63$\pm$0.04 & 133/147 \\
4 & 0.89 & 1.22$\pm$0.04 & 116/117 & 37 & 1.18 & 2.25$\pm$0.04 & 247/213 \\
5 & 0.97 & 1.33$\pm$0.04 & 129/128 & 38 & 0.58 & 3.43$\pm$0.08 & 204/186 \\
6 & 1.85 & 1.17$\pm$0.02 & 156/158 & 39 & 1.38 & 1.11$\pm$0.03 & 161/132 \\
7 & 1.51 & 1.21$\pm$0.02 & 161/154 & 40 & 0.90 & 1.20$\pm$0.03 & 135/121 \\
8 & 0.97 & 1.40$\pm$0.03 & 163/140 & 41 & 0.74 & 1.66$\pm$0.04 & 164/142 \\
9 & 1.27 & 1.30$\pm$0.03 & 152/148 & 42 & 0.76 & 1.41$\pm$0.03 & 133/127 \\
10 & 1.39 & 1.33$\pm$0.03 & 144/143 & 43 & 0.46 & 1.62$\pm$0.05 & 137/114 \\
11 & 0.40 & 2.01$\pm$0.08 & 93/108 & 44 & 0.48 & 1.78$\pm$0.05 & 132/121 \\
12 & 0.34 & 2.11$\pm$0.08 & 134/109 & 45 & 0.85 & 1.78$\pm$0.04 & 155/159 \\
13 & 1.51 & 1.09$\pm$0.02 & 145/141 & 46 & 1.08 & 2.59$\pm$0.04 & 208/218 \\
14 & 1.05 & 1.29$\pm$0.03 & 144/135 & 47 & 0.47 & 3.66$\pm$0.09 & 191/172 \\
15 & 0.74 & 1.45$\pm$0.04 & 130/128 & 48 & 0.87 &1.22$\pm$0.04 & 133/117 \\
16 & 0.55 & 1.65$\pm$0.05 & 110/120 & 49 & 1.05 & 1.14$\pm$0.03 & 134/123 \\
17 & 0.63 & 1.83$\pm$0.05 & 124/134 & 50 & 1.16 & 1.36$\pm$0.03 & 154/154 \\
18 & 1.15 & 1.48$\pm$0.03 & 162/152 & 51 & 1.03 & 1.34$\pm$0.03 & 154/147 \\
19 & 1.20 & 1.56$\pm$0.03 & 152/159 & 52 & 0.69 & 1.58$\pm$0.04 & 147/135 \\
20 & 0.14 & 5.38$\pm$0.27 & 145/115 & 53 & 0.71 & 1.81$\pm$0.04 & 140/146 \\
21 & 0.67 & 2.52$\pm$0.06 & 147/160 & 54 & 1.01 & 2.25$\pm$0.04 & 217/198 \\
22 & 1.38 & 1.13$\pm$0.02 & 165/136 & 55 & 1.32 & 2.16$\pm$0.03 & 214/205 \\
23 & 0.85 & 1.25$\pm$0.03 & 136/120 & 56 & 0.19 & 4.22$\pm$0.18 & 110/121 \\
24 & 0.62 & 1.42$\pm$0.04 & 117/116 & 57 & 1.15 & 1.33$\pm$0.03 & 163/144 \\
25 & 0.54 & 1.47$\pm$0.04 & 105/114 & 58 & 1.40 & 1.27$\pm$0.02 & 164/158 \\
26 & 0.58 & 1.58$\pm$0.04 & 122/125 & 59 & 1.36 & 1.37$\pm$0.03 & 157/153 \\
27 & 0.80 & 1.54$\pm$0.04 & 120/143 & 60 & 1.08 & 1.57$\pm$0.03 & 157/157 \\
28 & 1.21 & 1.58$\pm$0.03 & 162/167 & 61 & 1.04 & 1.74$\pm$0.04 & 127/155 \\
29 & 0.54 & 2.64$\pm$0.07 & 179/153 & 62 & 0.48 & 2.45$\pm$0.07 & 149/139 \\
30 & 1.09 & 1.55$\pm$0.03 & 188/151 & 63 & 0.49 & 1.71$\pm$0.06 & 130/115 \\
31 & 0.97 & 1.18$\pm$0.03 & 131/118 & 64 & 0.76 &1.60$\pm$0.04 & 126/133 \\
32 & 0.66 & 1.39$\pm$0.04 & 113/117 & 65 & 0.53 & 1.69$\pm$0.05 & 127/117 \\
33 & 0.64 & 1.37$\pm$0.04 & 135/116 & 66 & 0.09 & 4.63$\pm$0.31 & 101/94 
\enddata
\tablenotetext{a}{Error bars on the rolloff frequency $\nu_{\rm rolloff}$ represent the 90\% confidence range.}
\end{deluxetable*}

\begin{figure*}
\begin{center}
\includegraphics[width=\textwidth]{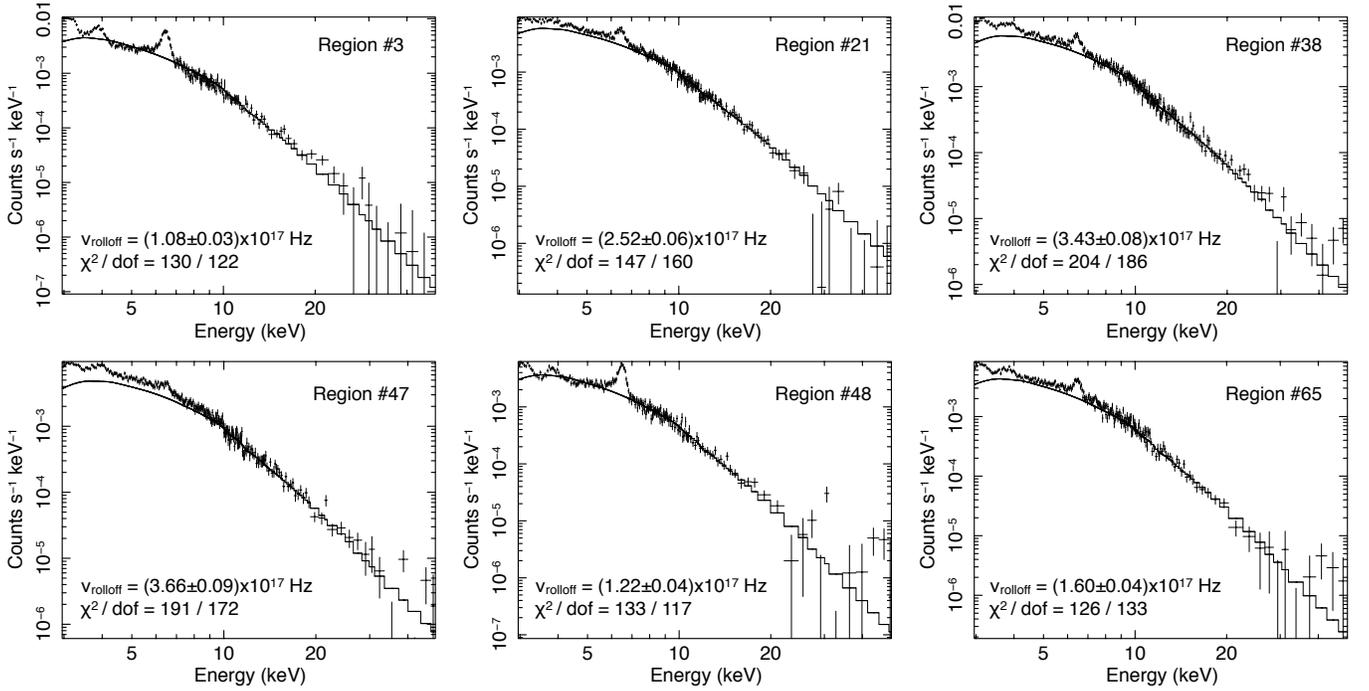}
\end{center}
\caption{Example spectra and fits using an absorbed {\it srcut} model for six regions. Although the spectra were fit over the range 10--50 keV where the non-thermal emission dominates, the spectra above span 3--50 keV to demonstrate the adequacy of the {\it srcut} model at predicting the non-thermal continuum across this energy range.}
\label{fig:spectra}
\end{figure*}

\begin{figure}
\begin{center}
\includegraphics[width=\columnwidth]{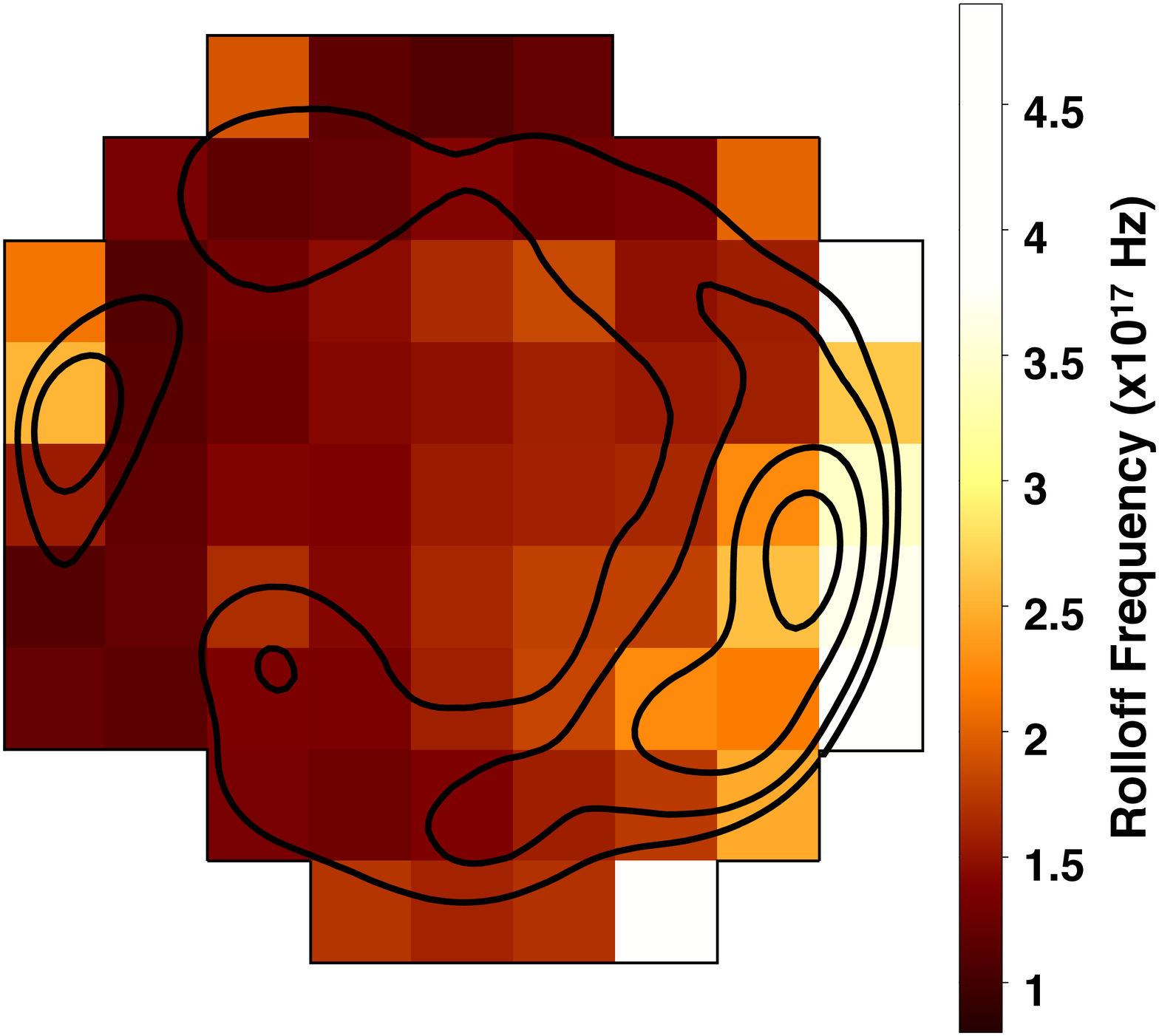}
\caption{Map of roll-off frequency $\nu_{\rm rolloff}$ across Tycho, with black contours from the 3--20 keV image in Figure~\ref{fig:boxes}. The softest spectra (i.e., those with the smallest best-fit values of $\nu_{\rm rolloff}$) are found in the southeast and north of Tycho, while the hardest spectra are in the northeast and the west. The hardest spectra (with $\nu_{\rm rolloff} \gs 3 \times 10^{17}$~Hz) are coincident with the ``stripes'' found in deep {\it Chandra} observations of Tycho by \cite{eriksen11}.}
\end{center}
\label{fig:srcutmaps}
\end{figure}

Table~\ref{tab:spec} lists the spectroscopic results, giving the best-fit rolloff frequency $\nu_{\rm rolloff}$ and the $\chi^{2}$ and degrees of freedom (d.o.f.) for all sixty-six regions. We find that all the regions are fit well by the {\it srcut} models, with $\chi^2$/d.o.f. = $\chi_{\rm r}^2\approx$0.82--1.26 with $\sim$100--200 degrees of freedom per region. Figure \ref{fig:spectra} shows example spectra from six of the regions, and Figure~\ref{fig:srcutmaps} maps the best-fit rolloff frequency $\nu_{\rm rolloff}$ values across Tycho. We find substantial variation of a factor of five in $\nu_{\rm rolloff}$. The largest values of $\nu_{\rm rolloff} \gs 3 \times 10^{17}$~Hz (corresponding to a rolloff energy $E_{\rm rolloff} = $1.24 keV) are concentrated in the west of the SNR where \cite{eriksen11} identified the ``stripes'' in Tycho. Furthermore, we find elevated rolloff frequencies in the northeast of the SNR, with $\nu_{\rm rolloff} \approx 2\times10^{17}$~Hz (corresponding to $E_{\rm rolloff} \approx$ 0.83 keV), where the SNR is thought to be interacting with molecular material \citep{reynoso99,lee04}. The lowest values of $\nu_{\rm rolloff} \sim 10^{17}$~Hz (corresponding to $E_{\rm rolloff} \approx 0.42$ keV) are found in the southeast and north of the SNR. For the range of $E_{\rm rolloff} \approx$0.4--2.0 keV we find in our spatially resolved spectroscopic analyses, Equation~\ref{eq:Emax} gives a maximum energy $E_{\rm max} \approx  (75-170) (B/\mu{\rm G})^{-1/2}$ TeV. Assuming a magnetic field $B=200 \mu$G across Tycho \citep{parizot06}, this relation gives $E_{\rm max} =$ 5--12 TeV. 

Previous X-ray studies of Tycho have also measured $\nu_{\rm rolloff}$ over the entire SNR or along prominent synchrotron filaments. Using {\it ASCA} data from Tycho, \cite{reynolds99} found $\nu_{\rm rolloff} \approx 8.8\times10^{16}$~Hz, and using the first {\it Chandra} observation of Tycho, \cite{hwang02} derived even lower $\nu_{\rm rolloff}$ toward the northeast, northwest, and southwest rim, with $\nu_{\rm rolloff} \sim$ 3--7 $\times10^{16}$~Hz. More recently, in their analysis of the deep {\it Chandra} observations of Tycho, \cite{eriksen11} fit absorbed {\it srcut} models to the hard ``stripes'' and a series of filaments projected in the south of Tycho (corresponding to our regions \#51 and 52) which they called the ``faint ensemble''. They found the stripes have $\nu_{\rm rolloff} \sim 2 \times 10^{18}$~Hz while the faint ensemble had softer spectra, with $\nu_{\rm rolloff} \approx 3 \times 10^{17}$~Hz. 

Our {\it NuSTAR} results confirm the stripes have harder emission than other regions in Tycho, but our 90\% confidence limits exclude $\nu_{\rm rolloff} \gs 5\times10^{17}$~Hz there. The softest regions (with $\nu_{\rm rolloff} \approx 10^{17}$~Hz) are comparable to the values found by \cite{reynolds99} integrated over Tycho. The discrepancy between our 90\% confidence limits and the hard stripes analyzed by \cite{eriksen11} may be due to two issues. First, the angular resolution of {\it NuSTAR} is not sufficient to resolve the individual stripes, and thus our spectra appear softer as they are averaging over larger areas of the SNR. Secondly, \cite{eriksen11} fit the {\it Chandra} data from 4.2--10 keV, and these spectra may not have had reliable leverage to model the high-energy rolloff. Additionally, we note that the effective area of {\it Chandra} drops precipitously above 8 keV. 

We note that we have assumed a constant radio spectral index of $\alpha=-0.65$, the value derived  by \cite{kothes06} based on the integrated flux densities at 408~MHz and 1420~MHz. However, the radio spectrum is actually concave, and thus, $\alpha$ may flatten at higher frequencies \citep{reynolds92}. If we instead adopted $\alpha=-0.5$ in our analysis, the best-fit values of $\nu_{\rm rolloff}$ change substantially, decreasing by about a factor of five. However, the assumption of $\alpha=-0.5$ is not favored statistically based on the $\chi^{2}$ computed from those fits. For example, region \#21 has $\chi^{2}=235$ (147) with 160 degrees of freedom in the $\alpha=-0.5$ ($-$0.65) case. Additionally, the $\alpha=-0.5$ fits dramatically over-predict the continuum flux below 10 keV, whereas the $\alpha=-0.65$ fits adequately describe the continuum down to 3 keV (see Figure~\ref{fig:spectra}). Generally, regardless of the choice of $\alpha$, the qualitative trend in the spatial distribution of $\nu_{\rm rolloff}$ (e.g., with greater $\nu_{\rm rolloff}$ in the west of the SNR) remains the same.

\section{Discussion} \label{sec:discuss}

\subsection{The Detection of $^{44}$Ti in Young SNRs} \label{sec:discuss_ti}

The yield, spatial distribution, and velocity distribution of $^{44}$Ti in a young SNR are a direct probe of the underlying explosion mechanism of the originating supernova (see e.g., \citealt{mag10}). Among core-collapse SNRs, Cassiopeia~A remains the only Galactic source with a confirmed detection of $^{44}$Ti \citep{iyudin94,brian14}, and SN~1987A in the Large Magellanic Cloud also has a robust $^{44}$Ti detection \citep{grebenev12,boggs15}. To date, the only probable detection of $^{44}$Ti in a Type Ia SNR was reported by \cite{troja14}, who found a 4-$\sigma$ level excess above the continuum in the 60--85 keV band in {\it Swift}/BAT observations of Tycho. \cite{wang14} also found a marginal 2.6-$\sigma$ detection of excess emission in the 60--90~keV band using {\it INTEGRAL}/IBIS, although their 3-$\sigma$ upper limit is consistent with the previous {\it INTEGRAL} studies of Tycho \citep{renaud06b}. Other searches for $^{44}$Ti in Type Ia SNRs have yielded upper limits, such as the recent work by \cite{zog14} using {\it NuSTAR} observations of the young SNR G1.9$+$0.3. 

Based on our {\it NuSTAR} analyses, we do not detect $^{44}$Ti in Tycho. We looked both for small scale ``clumpy" emission as found in Cassiopeia A \citep{brian14} as well as for diffuse emission throughout the SNR. We find upper limits which are more strict than the previous COMPTEL \citep{dupraz97,iyudin99} and {\it INTEGRAL} \citep{renaud06b,wang14} values, particularly if the $^{44}$Ti is concentrated in the central 2\arcmin\ and/or has slow or moderate velocities (i.e., the 0.1~keV and 1.0~keV line width cases). For example, our 3-$\sigma$ limit in the 0.1~keV (1.0~keV) case at an enclosed radius of 2\arcmin\ is 5.0$\times10^{-6}$ (1.0$\times10^{-5}$) ph cm$^{-2}$ s$^{-1}$. To be consistent with the possible detection of $^{44}$Ti with {\it Swift}/BAT \citep{troja14}, our results indicate that the $^{44}$Ti is expanding at moderate-to-high velocities ($\gs$7000 km s$^{-1}$) and/or is distributed across the majority ($\gs$3\arcmin) of the SNR.

We have also found no $^{44}$Ti in the regions where \cite{miceli15} reported detection of the Ti-K line complex using {\it XMM-Newton}. This $\sim$4.9~keV feature arises from shock-heated, stable $^{48}$Ti and $^{50}$Ti. Thus, the lack of a co-located detection of $^{44}$Ti may indicate that the radioactive isotope does not trace the Fe-rich ejecta. Alternatively, the divergent results may suggest that the 4.9~keV feature arises from transitions of Ca {\sc xix} and Ca {\sc xx} rather than from stable Ti. In the future, these interpretations can be tested via deeper {\it NuSTAR} observations to reveal the spatial distribution of $^{44}$Ti, and through high-resolution X-ray spectroscopy (with e.g., {\it Astro-H}) which may distinguish between the Ca and Ti spectral features. 

The upper limit flux we have measured of the 67.9~keV line can be employed to set an upper limit on the $^{44}$Ti yield in the SN explosion. The $^{44}$Ti mass $M_{44}$ is related to the flux $F_{68}$ in the 67.9~keV $^{44}$Ti line (e.g., \citealt{grebenev12}) by

\begin{equation}
M_{44} = 4 \pi d^2 (44 m_{\rm p})~\tau~{\rm exp}(t/\tau)~F_{68}~W^{-1},
\end{equation}

\noindent
where $d$ is the distance to the source, $m_{\rm p}$ is the mass of a proton, $\tau$ is the mean lifetime of the radioactive titanium (which is related to the half-life $t_{1/2}$ by  $\tau = t_{1/2} / {\rm ln}~2 \approx$~85~years for $^{44}$Ti: \citealt{ahmad06}), $t$ is the age of the source, and $W$ is the emission efficiency (average numbers of photons per decay) for a given emission line ($W =$ 0.93 for the 67.9~keV $^{44}$Ti line). Assuming a distance $D=2.3$~kpc and an age of $t =$~442~years for Tycho, an upper limit of $F_{68} < 2 \times 10^{-5}$~ph~cm$^{-2}$~s$^{-1}$ corresponds to $M_{44} < 2.4 \times 10^{-4} M_{\sun}$. Here, we adopted the upper limit $F_{68}$ in the case where the $^{44}$Ti is located within a radius of 3\arcmin\ and has a line width of 1~keV, as these most closely match the expected scale and velocity of the shell. For the range of distance estimates in the literature of $D = 1.7$--5.0 kpc to Tycho, we find upper limits of $M_{44} < 1.3 \times 10^{-4} M_{\sun}$ and $M_{44} < 10^{-3} M_{\sun}$, respectively. 

These upper limits can be compared to the predictions of $^{44}$Ti synthesized in Type Ia SN models. Generally, the amount of $^{44}$Ti produced in a Type Ia SN depends on both the progenitor systems as well as the ignition processes. Simulations predict $^{44}$Ti masses ranging from $\sim$10$^{-6} M_{\sun}$ in centrally ignited pure deflagration models and $\ls6 \times 10^{-5} M_{\sun}$ for off-center delayed detonation models (e.g., \citealt{iwamoto99,maeda10}). In sub-Chandrasekhar models, greater $^{44}$Ti yields are expected, with estimates of $\gs$10$^{-3} M_{\sun}$  (e.g., \citealt{fink10,woosley11,moll13}). The mass limits estimated above are most consistent with the models that produce small or moderate yields of $^{44}$Ti and disfavor the sub-Chandrasekhar models.

\subsection{Particle Acceleration in Tycho} \label{sec:discuss_pa}

To explore the relationship between the non-thermal X-ray emission detected by {\it NuSTAR} and shock properties, we compare the spatially resolved spectroscopic results of Section~\ref{sec:spectra} to measurements of the shock expansion and the ambient density $n_{\rm o}$ in Tycho from previous studies. 

Expansion measurements of Tycho's rim have been done at radio, optical, and X-ray wavelengths. Using 1375-MHz VLA data over a 10-year baseline, \cite{reynoso97} measured expansion rates of $\sim$0.1--0.5\arcsec\ yr$^{-1}$, depending on the azimuthal angle. These authors found the slowest expansion occurs in the east/northeast, where the forward shock is interacting with a dense cloud \citep{reynoso99,lee04}, and the fastest expansion is along the western rim. Quantitatively similar results were obtained in the optical by \cite{kamper78} and in the X-ray by \cite{katsuda10}. Using the measured expansion index and SNR evolutionary models, \cite{katsuda10} inferred a pre-shock ambient density of $n_{\rm o} \ls$ 0.2 cm$^{-3}$, with probable local variations of factors of a few based on the substantial differences in the proper motion around the rim.

The shock expansion is described by the radius of the shock with time, $R \propto t^{\rm m}$, where $m$ is the expansion parameter. From this expression, the shock velocity is related to $m$ by $v_{\rm s} = mR/t$. Thus, the $m$ is a useful indicator of shock velocity, yet it does not require assumptions about the distance to the source. Generally, $m$ is close to unity when the expansion is undecelerated (as in e.g., the free expansion phase of a young SNR). As the shock undergoes deceleration, $m \sim$0.6--0.8 \citep{chev82b,chev82a}, and $m=0.4$ once the shock reaches the Sedov phase. Finally, as the shock becomes radiative, $m$ drops below 0.4. While interior pressure is still important, $m=0.33$ (``pressure-driven snowplow"; e.g. \citealt{blondin98}), and then falls to $m=0.25$ when the shock transitions to a pure momentum-conserving stage \citep{cioffi88}. In the case of Tycho, \cite{reynoso97} demonstrated the expansion parameter $m$ varies around the rim from $m\sim$ 0.25 in the east (where the shock is decelerating due to the collision with a dense clump known as {\it knot g}: \citealt{kamper78,ghav00}) up to $m \ls$0.8 in the southwest.

\begin{figure}
\begin{center}
\includegraphics[width=\columnwidth]{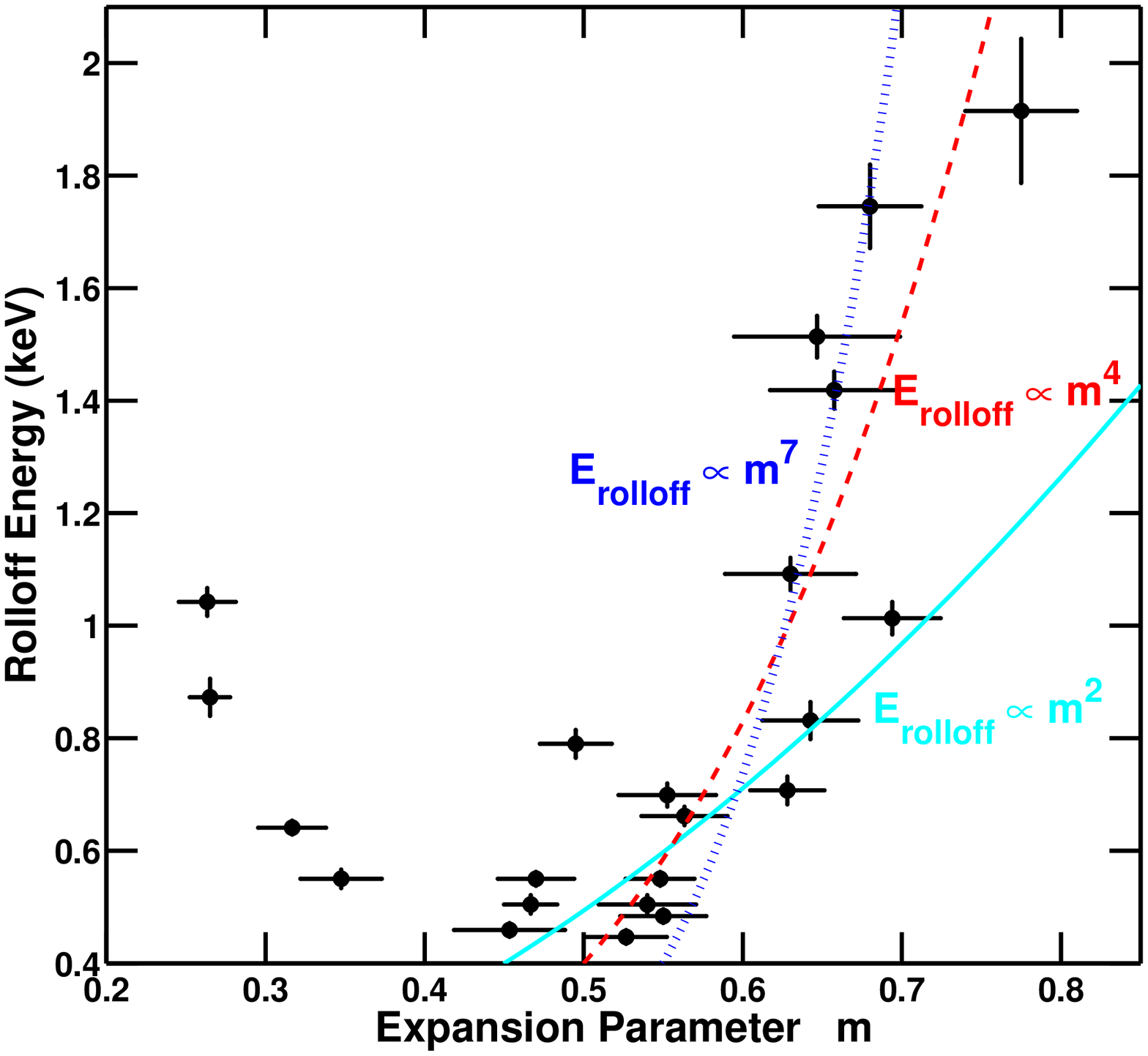}
\end{center}
\caption{Rolloff energy $E_{\rm rolloff}$ (determined in Section~\ref{sec:spectra}) of regions around the rim of Tycho versus expansion parameter $m$ (as listed in Table~2 of \citealt{reynoso97}). Overplotted are three curves which represent different dependences of $E_{\rm rolloff}$ on $m$: $E_{\rm rolloff} \propto m^2$ (cyan solid line), $E_{\rm rolloff} \propto m^4$ (red dashed line), and $E_{\rm rolloff} \propto m^7$ (blue dotted line). These dependences are derived using the expressions in Equations~\ref{eq:agelimited}--\ref{eq:esclimited} and given that $R \propto t^{m}$, so $v_{\rm s} = mR/t \propto m$. The former two trends are those expected if the maximum energy of the electrons is limited by the age of the SNR or by the timescale of the radiative losses, respectively. The latter curve represents the loss-limited case if the B-field strength is not constant (and instead $B \propto v_{\rm s} \propto m$). The $E_{\rm rolloff} \propto m^2$ trend does not match the data for large $m$, and thus, the age-limited scenario is more consistent with our results. }
\label{fig:Evsm}
\end{figure}

In Figure~\ref{fig:Evsm}, we show the rolloff energy (found in Section~\ref{sec:spectra}) versus the expansion parameter $m$ obtained by \cite{reynoso97} using VLA data. To produce this plot, we have considered only our regions around the rim of Tycho, and we have adopted the mean of the expansion parameters over the azimuthal angles which correspond to those regions using the data in Table~2 of \citealt{reynoso97}. At $m>0.4$, the rolloff energy appears to increase with $m$, suggesting the least decelerated (fastest) shocks of Tycho are the most efficient at accelerating particles to high energies. The minimum rolloff energy is around $m=0.4$ where the SNR has reached its Sedov phase. At $m = 0.25$ where the shock is encountering dense material, the SNR is slightly more efficient at accelerating particles, reaching half of the rolloff energy of the high-$m$ regions. 

\begin{figure*}
\begin{center}
\includegraphics[width=0.48\textwidth]{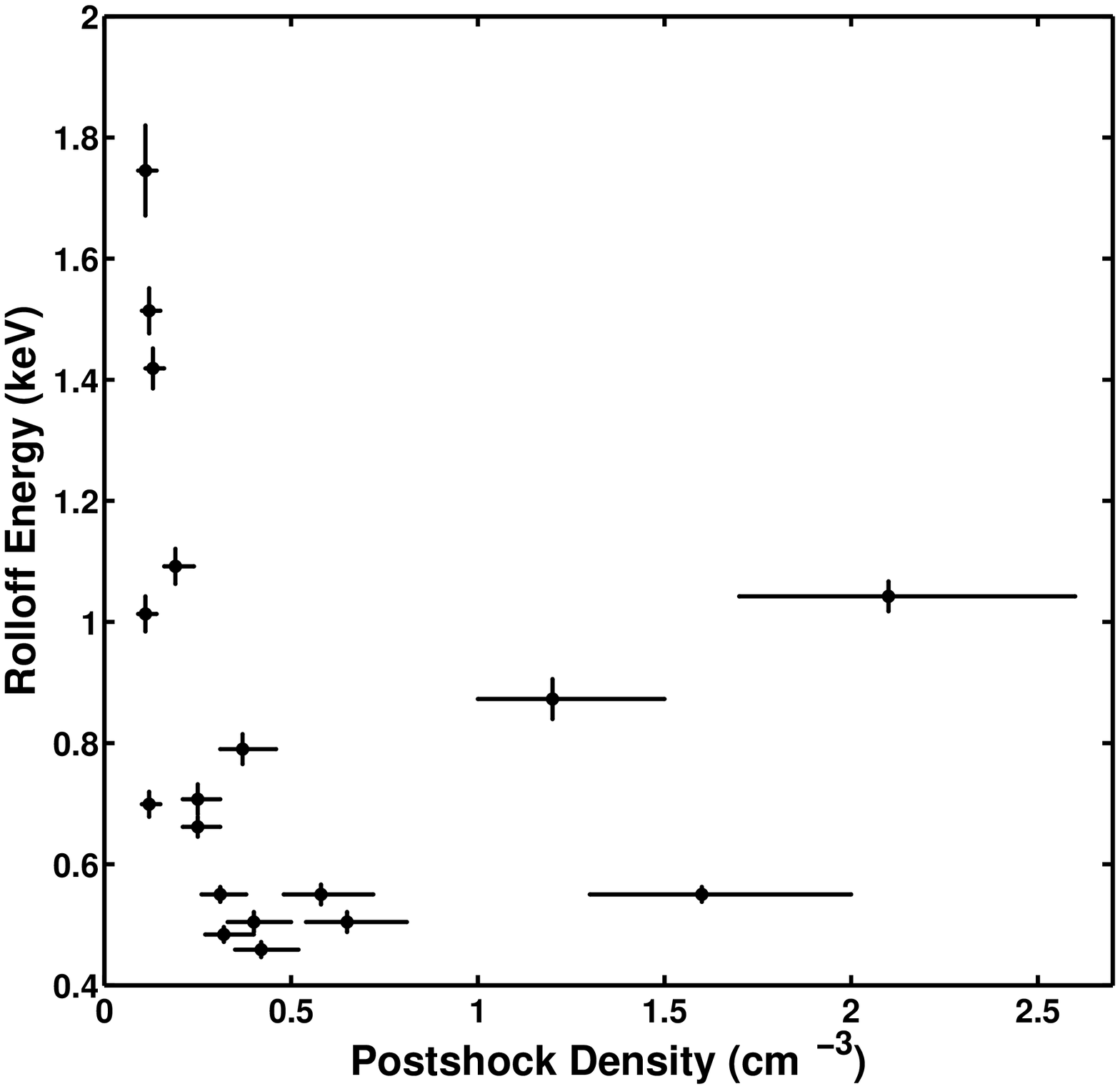}
\includegraphics[width=0.48\textwidth]{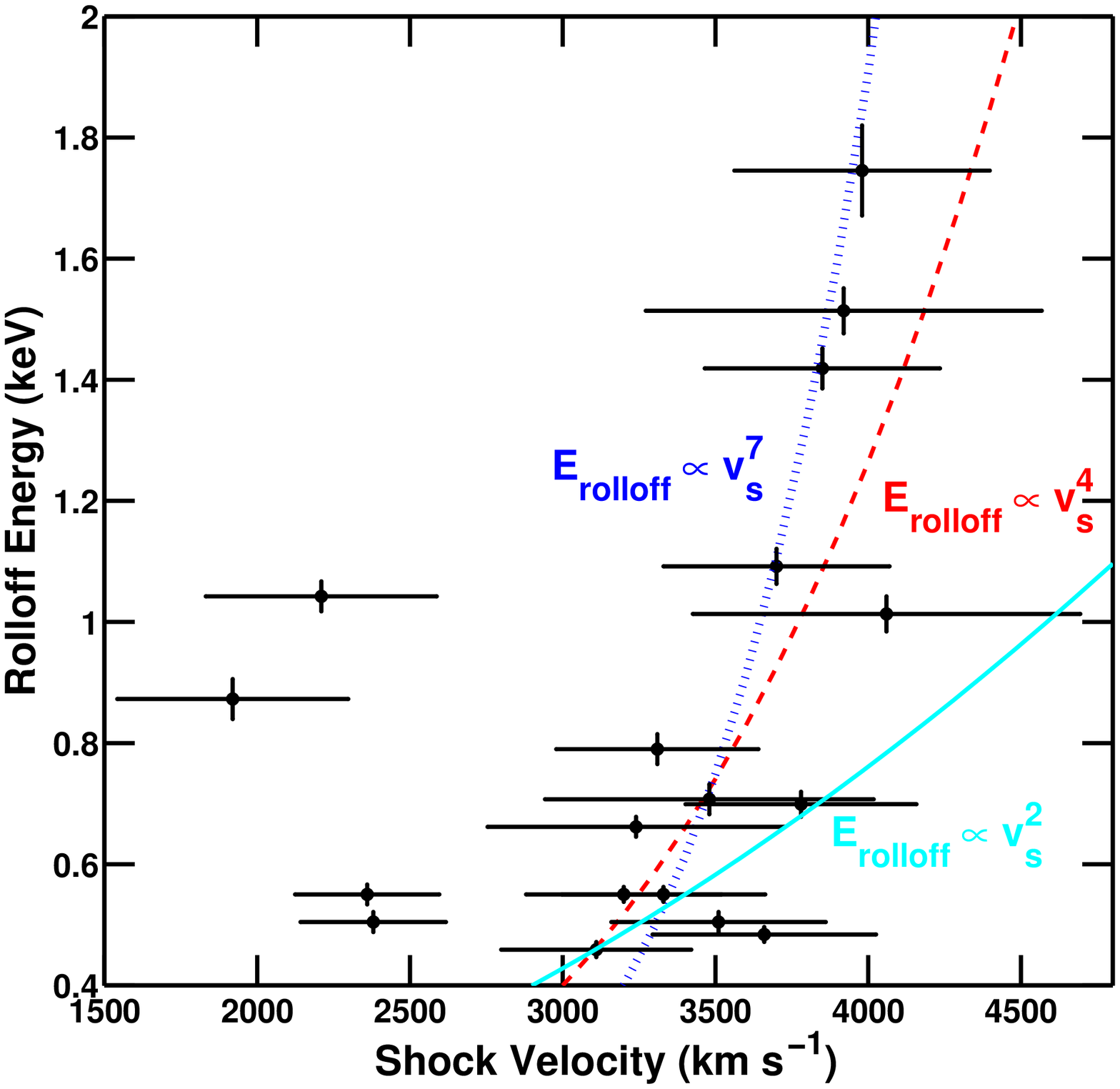}
\end{center}
\caption{Rolloff energy $E_{\rm rolloff}$ versus post-shock density $n_{\rm o}$ (left) and shock velocity $v_{\rm s}$ (right) derived by \cite{williams13} adopting a distance of $D=$ 2.3 kpc. The velocities scale linearly with the assumed distance, but the trend of $E_{\rm rolloff}$ with $v_{\rm s}$ is independent of distance. Overplotted on the right panel are three curves which represent different dependences of $E_{\rm rolloff}$ on shock velocity $v_{\rm s}$: $E_{\rm rolloff} \propto v_{\rm s}^2$ (cyan solid line), $E_{\rm rolloff} \propto v_{\rm s}^4$ (red dashed line), and $E_{\rm rolloff} \propto v_{\rm s}^7$ (blue dotted line). The former two trends are those expected if the maximum energy of the electrons is limited by the age of the SNR or by the timescale of the radiative losses, respectively. The latter curve represents the loss-limited case if the B-field strength is not constant (and instead $B \propto v_{\rm s}$). The $E_{\rm rolloff} \propto v_{\rm s}^2$ trend does not match the data, and the age-limited scenario is more consistent with our results. We note that the three $n_{0} >$1~cm$^{-3}$ points correspond to regions \#12, 10, and 21 (in order of increasing $n_{0}$). The two $v_{\rm s} \ls 2200$~km~s$^{-1}$ points correspond also to regions \#12 and 21, which are located where the forward shock is interacting with the dense clump know as {\it knot g} \citep{kamper78,ghav00}.}
\label{fig:Evsbw}
\end{figure*}

In Figure~\ref{fig:Evsbw}, we plot the rolloff energy $E_{\rm rolloff}$ versus the post-shock density $n_{\rm o}$ (left panel) and versus the shock velocity $v_{\rm s}$ (right panel). Both of these quantities were derived by \cite{williams13} using {\it Spitzer Space Telescope} 24 and 70 $\mu$m data as well as the radio and X-ray proper motions of \cite{reynoso97} and \cite{katsuda10}. We find that the regions of large $E_{\rm rolloff}$ have the lowest post-shock densities (with $n_{\rm o} \ls 0.25$ cm$^{-3}$) and the largest shock velocities (with $v_{\rm s} \gs$ 3500 km s$^{-1}$). The two outliers to these trends (those with $n_{0} >$1 cm$^{-3}$, $v_{\rm s} <$ 2200 km s$^{-1}$, and $E_{\rm rolloff} >$0.8~keV) correspond to regions \#12 and 21, where the shock is interacting with {\it knot g}. 

We note that the conversion of proper motions to shock velocities requires an assumption about the distance to Tycho, which is still uncertain, with estimates in the literature ranging from 1.7--5.0 kpc (e.g., \citealt{albinson86,schwarz95,volk08,hayato10,tian11,slane14}). For the data plotted in Figure~\ref{fig:Evsbw}, \cite{williams13} employed a distance of $D=$ 2.3 kpc, as suggested by \cite{chev80}. The shock velocities scale linearly with the assumed distance, so the choice of distance will shift the points in the right panel of Figure~\ref{fig:Evsbw} accordingly.

The relationship between $E_{\rm rolloff}$ and $v_{\rm s}$ in Figure~\ref{fig:Evsbw} can be used to examine the mechanism limiting the maximum energy of electrons $E_{\rm max}$ undergoing diffusive shock acceleration. The timescale $t_{\rm acc}$ of the acceleration is set by the lesser of three relevant processes: the age of the SNR (i.e., the finite time the electrons have had to accelerate), the timescale of the radiative losses of the electrons, and the timescale for the particles to escape upstream of the shock. In each case, $E_{\rm max}$ (and consequently, $E_{\rm rolloff}$) has different dependences on physical parameters (from \citealt{reynolds08}, Equations 26--28), specifically $v_{\rm s}$:

\begin{equation}
E_{\rm max} ({\rm age}) \propto v_{\rm s}^2 t B (\eta R_{\mathcal{J}})^{-1} 
\end{equation}

\begin{equation}
E_{\rm max} ({\rm loss}) \propto v_{\rm s} ( \eta R_{\mathcal{J}} B)^{-1/2}
\end{equation}

\begin{equation}
E_{\rm max} ({\rm esc}) \propto B \lambda_{\rm max}.
\end{equation}

\noindent
In the above relations, $t$ is the age of the SNR, $\eta$ is the gyrofactor (where $\eta=1$ is Bohm diffusion and $\eta >$1 is ``Bohm-like'' diffusion), and $\lambda_{\rm max}$ is the maximum wavelength of MHD waves present to scatter electrons. $R_{\mathcal{J}}$ parametrizes the obliquity dependence of the acceleration: given that the shock obliquity angle $\theta_{\rm Bn}$ is the angle between the mean upstream B-field direction and the shock velocity, then the quantity $R_{\mathcal{J}}$ is defined as $R_{\mathcal{J}} \equiv t_{\rm acc}(\theta_{\rm Bn})/t_{\rm acc}(\theta_{\rm Bn}=0).$ Since $E_{\rm rolloff} \propto E_{\rm max}^2 B$, the above relations can be rewritten in terms of $E_{\rm rolloff}$: 

\begin{equation}
E_{\rm rolloff} ({\rm age}) \propto v_{\rm s}^4 t^2 B^3 (\eta R_{\mathcal{J}})^{-2} 
\label{eq:agelimited}
\end{equation}

\begin{equation}
E_{\rm rolloff} ({\rm loss}) \propto v_{\rm s}^2 ( \eta R_{\mathcal{J}})^{-1}
\end{equation}

\begin{equation}
E_{\rm rolloff} ({\rm esc}) \propto B^3 \lambda_{\rm max}^2.
\label{eq:esclimited}
\end{equation}

In the right panel of Figure~\ref{fig:Evsbw}, we plot curves for $E_{\rm rolloff} \propto v_{\rm s}^2$ (solid light blue line) and $E_{\rm rolloff} \propto v_{\rm s}^4$ (dashed red line), for comparison. Additionally, since $v_{\rm s} = mR/t \propto m$ (where $R$ is the radius of the shock and $t$ is the age of the SNR), we have plotted the same curves on Figure~\ref{fig:Evsm} as well. The two dependencies of $E_{\rm rolloff}$ on $v_{\rm s}$ reflect those of the loss-limited and age-limited relations above, assuming constant $B$ and $R_{\mathcal{J}}$. We find that the $\propto v_{\rm s}^{2}$ curve is too shallow to match the high-velocity ($ v_{\rm s} \gs$ 3500 km s$^{-1}$) points, and the $\propto v_{\rm s}^4$ curve is a better descriptor of the data. We note that if $B$ is not constant and is instead amplified by a constant fraction of the shock kinetic energy, then $B \propto v_{\rm s}$, and Equation~\ref{eq:agelimited} gives $E_{\rm rolloff} \propto v_{\rm s}^7$. We have plotted this steeper curve in Figures~\ref{fig:Evsm} and ~\ref{fig:Evsbw} as well, and it adequately describes the data. Due to the relatively large error bars on the shock velocity $v_{\rm s}$, we are not able to distinguish which curve ($E_{\rm rolloff} \propto v_{\rm s}^4$ or $E_{\rm rolloff} \propto v_{\rm s}^7$) is better statistically, but both are much more successful than the loss-limited case (with $E_{\rm rolloff} \propto v_{\rm s}^2$, with no dependence on magnetic-field strength $B$) and the escape-limited case (with no $E_{\rm rolloff}$ dependence on $v_{\rm s}$). Thus, our spatially resolved spectroscopic results are most consistent with the age-limited scenario, whereby the maximum electron energy $E_{\rm max}$ is determined by the finite time the electrons have had to accelerate. In this case, the accelerated electrons achieve the same maximum energy as the protons/ions because the electrons are not limited by their radiative losses.

However, previous studies of Tycho have argued the SNR is likely loss-limited instead of age-limited (although see \citealt{reynolds99}). In particular, \cite{parizot06} and \cite{helder12} calculate the synchrotron loss timescale and found it to be a few percent of the age of Tycho. Furthermore, recent hydrodynamical simulations by \cite{slane14} show the loss-limited scenario is consistent with the observed broadband spectrum of Tycho. We note that these models were matched to the global spectra across Tycho as the gamma-ray emission is not resolved. Thus, local variations in the gamma-ray emission is possible, with the implication that the electrons and protons may achieve the same maximum energies in certain locations. 

For the acceleration to be age limited, the synchrotron loss time $t_{\rm syn}$ for X-ray emitting electrons must be larger than the remnant age, $t=$442~years for Tycho. As a consistency check, we estimate the magnetic field strength $B$ necessary for $t_{\rm syn} > t$ and compare it to that measured observationally. The synchrotron loss time $t_{\rm syn}$ is related to the $E_{\rm max}$ and $E_{\rm rolloff}$ by (adapted from \citealt{vink12})

\begin{eqnarray}
t_{\rm syn} & = & 12.5 \bigg( \frac{E_{\rm max}}{100~{\rm TeV}} \bigg)^{-1} \bigg( \frac{B}{100~\mu{\rm G}} \bigg)^{-2}~{\rm yr} \\
t_{\rm syn} & = & 104 \bigg( \frac{E_{\rm rolloff}}{1~{\rm keV}} \bigg)^{-1/2} \bigg( \frac{B}{100~\mu{\rm G}} \bigg)^{-3/2}~{\rm yr}
\end{eqnarray}

\noindent
Adopting the highest rolloff frequency from our spectral analysis, $\nu_{\rm rolloff} = 5.38 \times 10^{17}$~Hz (from region \#20), we derive $B = 29~\mu{\rm G}$. This value is far below the measurements by \cite{parizot06} of $B=200 \mu$G based on the assumption that the X-ray filament widths are limited by the synchrotron loss time. However, it is possible that filament widths are governed instead by magnetic field damping (e.g., \citealt{pohl05,rettig12,ressler14}), and this mechanism is the only possible one if the rims are equally thin at radio wavelengths, as is true in some regions of Tycho. Thus, the age-limited interpretation of our {\it NuSTAR} results would demand low magnetic field strengths and magnetic-field damping to be responsible for the X-ray rim morphology. However, a recent detailed analysis of Tycho \citep{tran15} finds that even in damping models, magnetic field strengths $\gg$30 $\mu$G are required to fit the radio and X-ray profiles of the thin rims.

Alternatively, the electrons are loss limited, and the apparent trend of $E_{\rm rolloff}$ with $v_{\rm s}$ arises because of obliquity effects. In the scalings relating $E_{\rm rolloff}$ to $v_{\rm s}$, we have assumed constant $R_{\mathcal{J}}$, yet the shock obliquity $\theta_{\rm Bn}$ is expected to change around the periphery of the SNR as the shock encounters a roughly uniform magnetic field in the environs of a Type Ia SN \citep{reynolds98}. For example, changes in obliquity are thought to be the origin of the X-ray bright, synchrotron rims aligned in the northeast and southwest directions of SN~1006 \citep{rothenflug04,reynoso13}. The magnetic field is not as ordered in Tycho as in SN~1006 (see Figure~8 of \citealt{reynoso97}), and further work is necessary to assess the obliquity dependence of the particle acceleration in Tycho. 

\section{Conclusions} \label{sec:conclusions}

We have reported the large, $\sim$750~ks {\it NuSTAR} observing program toward the historical supernova remnant Tycho. Using these data, we produced narrow-band images over several energy bands to identify the locations of the hardest X-rays and to search for radioactive decay line emission from $^{44}$Ti. We find that the hardest $>$10 keV X-rays are concentrated to the southwest of Tycho, where recent {\it Chandra} observations have revealed high emissivity ``stripes'' associated with particles accelerated to the knee of the cosmic-ray spectrum. We do not find evidence of $^{44}$Ti, and we set limits on the presence of $^{44}$Ti, depending on the velocity and distribution of the metal. In order to be consistent with the reported {\it Swift}/BAT detection, the Ti must be expanding at moderate-to-high velocities and/or distributed over the majority of the SNR. Our spatially-resolved spectroscopic analyses of sixty-six regions showed that the highest energy electrons are accelerated by the fastest shocks and in the lowest density regions of the SNR. We find a steep dependence of the roll-off frequency with shock velocity that is consistent with the maximum energy of accelerated electrons being limited by the age of the SNR rather than by synchrotron losses, contrary to previous results obtained for Tycho. One way to reconcile these discrepant findings is through shock obliquity effects, and future observational work is necessary to explore the role of obliquity in the particle acceleration process.

\acknowledgements

We acknowledge helpful discussions with Drs. Marco Miceli, Lorenzo Sironi and Patrick Slane. LAL received support for this work from NASA through Hubble Fellowship grant number HST-HF2-51342.001 awarded by the Space Telescope Science Institute, which is operated by the Association of Universities for Research in Astronomy, Inc., for NASA, under contract NAS 5--26555. Additionally, the work was supported under NASA contract NNG08FD60C and made use of data from the {\it NuSTAR} mission, a project led by the California Institute of Technology, managed by the Jet Propulsion Laboratory, and funded by NASA. We thank the {\it NuSTAR} Operations, Software and Calibration teams for support with the execution and analysis of these observations. This research made use of the {\it NuSTAR} Data Analysis Software (NuSTARDAS), jointly developed by the ASI Science Data Center (ASDC, Italy) and the California Institute of Technology (USA). \\

\noindent
{\it Facilities}: {\it NuSTAR}, {\it Chandra}, VLA

\nocite{*}
\bibliographystyle{apj}
\bibliography{tycho}

\end{document}